\documentclass{Interspeech2024}
\usepackage{environ}
\NewEnviron{FitToWidth}[1][\textwidth]{%
\begin{center}
  \makebox[#1]{\resizebox{#1}{!}{\BODY}}%
\end{center}
}
\usepackage{tipa}

\interspeechcameraready

\title{Enhancing Child Vocalization Classification with Phonetically-Tuned Embeddings for Assisting Autism Diagnosis}

\name[affiliation={1,2}]{Jialu}{Li}
\name[affiliation={1,2}]{Mark}{Hasegawa-Johnson}
\name[affiliation={3}]{Karrie}{Karahalios}

\address{
  $^1$Department of Electrical and Computer Engineering, University of Illinois\\
  $^2$Beckman Institute for Advanced Science and Technology, University of Illinois \\
  $^3$Department of Computer Science, University of Illinois}
\email{jialuli3@illinois.edu, jhasegaw@illinois.edu, kkarahal@illinois.edu}


\begin{document}

\maketitle

\begin{abstract}
    
The assessment of children at risk of autism typically involves a clinician observing, taking notes, and rating children’s behaviors. A machine learning model that can label adult and child audio may largely save labor in coding children's behaviors, helping clinicians capture critical events and better communicate with parents. In this study, we leverage Wav2Vec 2.0 (W2V2), pre-trained on 4300-hour of home audio of children under 5 years old, to build a unified system for tasks of clinician-child speaker diarization and vocalization classification (VC). To enhance children’s VC, we build a W2V2 phoneme recognition system for children under 4 years old, and we incorporate its phonetically-tuned embeddings as auxiliary features or recognize pseudo phonetic transcripts as an auxiliary task. We test our method on two corpora (Rapid-ABC and BabbleCor) and obtain consistent improvements. Additionally, we outperform the state-of-the-art performance on the reproducible subset of BabbleCor.
\end{abstract}

\section{Introduction}

Autism often manifests in early childhood in the form of social communication deviations and the presence of restricted interests and repetitive behaviors~\cite{hodges2020autism}. According to estimates from the Centers for Disease Control and Prevention, 1 in 36 children has been identified with autism in 2020 in the U.S.~\cite {maenner2023prevalence}. Early autism detection can improve life outcomes by encouraging an autism-friendly environment~\cite{fernell2013early}, but acquiring autism diagnostic outcomes through conventional evaluations may be protracted due to the limited available healthcare professional services and the complexity of examination procedures~\cite{gordon2016whittling}. Hence, a robust machine learning (ML) model, capable of automatically detecting features that clinicians consider relevant to the early diagnosis,  may be helpful for clinicians to capture events of interest and better explain conversation patterns indicative of autism to parents. 

Previous work has built ML-based child-adult speaker diarization (SD) models for Autism Diagnostic Observation Schedule
(ADOS) interviews~\cite{lord2000autism,9053251,krishnamachari2021developing}. ADOS is a well-established protocol administrated by a trained clinician to elicit children's speech. 
Samples of ADOS interviews between clinicians and children aged 12-16 years were used in the series of DIHARD SD challenges~\cite{ryant2020third}.

To help diagnose toddlers aged 1-2, who are in the early stages of language development and often can't express themselves through complete sentences, it's crucial for the ML model to perform children's vocalization classification (VC), in addition to the child-adult SD task. Autism researchers have reported that a low frequency of verbalization and non-verbal cues (vocalization, laughter, crying, etc.) is one of the early behavioral signs for autistic children~\cite{ozonoff2010prospective}. Past studies have analyzed the acoustic differences between typical developing children and those with autism~\cite{lyakso2016comparison,mohanta2022analysis}, and have also explored using speech-related features to perform automatic classification between these two groups~\cite {cho2019automatic,santos2013very}. Baird et al. \cite{baird2017automatic} explored using a range of speech-based classification methods to assess levels of severity of autism, while Cheng et al.~\cite{cheng2023computer} classified autism on the basis of pre-defined behavioral templates. Beyond classification of autism, past work also studied preschool-aged child VC for a robot-assisted diagnostic protocol~\cite{8443030}.

Autism researchers have relied on manually coding children's social behaviors for assessment, but coding is a time- and labor-intensive task. To combat data sparsity issues, self-supervised learning models, such as wav2vec 2.0 (W2V2) ~\cite{wav2vec}, have performed well on several speech processing downstream tasks
given a limited amount of labeled data. W2V2 first uses large-scale unlabeled data for pre-training followed by fine-tuning on a small amount of labeled data. A few recent studies also explored using self-supervised learning models for child-adult SD and VC tasks~\cite{xu2023understanding,lahiri2023robust}. 

In this study, we explore building a unified end-to-end W2V2-based system for SD and VC tasks on clinician-child recordings for autism assessment using the Rapid-ABC (RABC) corpus. We propose to incorporate phonetically-tuned embeddings, learned from a W2V2-based phone recognizer (W2V2-PR) for children under 4 years old, to improve children's VC, despite the imperfect accuracy of the phone recognizer. Specifically, We explore three methods including (1) using W2V2-PR representations as auxiliary input features, (2) recognizing pseudo phonetic transcripts generated from the W2V2-PR system as an auxiliary output task, and (3) combining (1) and (2). 
To the best of our knowledge, this is the first study that leverages a self-supervised phone recognizer for children under 4 years old to enhance the child VC task.
We further demonstrate the superiority of our proposed method in children's VC on two corpora that shared similar annotation protocols (RABC and BabbleCor).

\section{Data}
In this study, we experiment with RABC for child-adult SD and VC tasks, and we further validate our method of improving children's VC results on BabbleCor. We use MyST and Providence for building children's phone recognition.

\noindent\textbf{Rapid-ABC (RABC)}~\cite{Rehg_2013_CVPR} contains both video and audio recordings of 3-5 mins brief interactive assessment between a child of 1-2 years old and a clinician. Two audio streams are recorded separately, each from a lapel mic placed in front of a speaker's chest. 
Two types of annotations are available: protocol segment, and child vocalization.  Protocol segment annotations mark the times at which the adult first speaks each of five key phrases, marking the start of five parts of the diagnostic protocol, including greetings, playing with the ball, reading the book, putting the book over the head, and smiling and tickling. Child vocalization carefully labels child audio as one of non-lexical vocalization (\texttt{VOC}), verbalization that contains words (\texttt{VERB}), cry (\texttt{CRY}), or laugh (\texttt{LAU}). To evaluate SD, we manually label adult audio as one of \texttt{VOC} or \texttt{LAU}. Ten percent of adult audio was double-coded, and inter-coder reliability (Cohen’s kappa score) was $\kappa=0.90$ at a precision of 0.2s. 
In this study, we analyze 51 sessions of audio recordings from part of the RABC corpus available for research use, where a total of 4 clinicians and 43 children participated (with 8 children being assessed twice).
To prevent over-fitting, we use 3-fold cross-validation for evaluation. No child is ever part of both the training and testing sets in any fold. Table~\ref{tab:rabc_data} presents partition details for each fold. For fine-tuning W2V2, we label the audio stream in frames of 2s starting every 0.1s; the label is determined by the central 0.1s of each frame. The mean and standard deviation (std) of the child utterances is 1.54s $\pm$ 2.30s.

\begin{table}
  \centering
     \caption{
   Number of participants (``\# of ID'') and duration of audio (``m.''=minutes) in each of six categories in each RABC cross-validation fold.  Symbol ``/'' separates training and testing durations. ``Total time'' counts the length of entire recordings, including silence and non-speech events.}
\setlength{\tabcolsep}{2.0pt}
  \begin{FitToWidth}[\columnwidth]
  \begin{tabular}{c|c|cc|cccc|c}
    \toprule
    \multicolumn{1}{c|}{}& \multicolumn{1}{c|}{}& \multicolumn{2}{|c|}{\texttt{ADU} m.} & \multicolumn{4}{c}{\texttt{CHI} m.} & \multicolumn{1}{|c}{Total}\\
    \multicolumn{1}{c|}{Fold \# } & \multicolumn{1}{c}{\# of ID} &  \multicolumn{1}{|c}{\texttt{VOC}} & \texttt{LAU} & \texttt{VOC} & \texttt{VERB} & \texttt{LAU} & \texttt{CRY} & \multicolumn{1}{c}{Time m.} \\
    \midrule
    1& 28 / 15& 38.6/20.3 & 2.4/1.3 & 4.4/1.4 & 2.3/1.5 & 1.4/.38 & 7.0/.53 & 93.9/48.3 \\
    \midrule
    2& 29 / 14& 39.4/19.5 & 2.7/1.0 & 4.0/1.8 & 2.3/1.6 & 1.1/.73 & 1.7/5.9 & 93.9/48.4 \\
    \midrule
    3& 29 / 14& 39.9/19.1 & 2.3/1.4 & 3.1/2.6 & 3.1/.77 & 1.1/.68 & 6.5/1.1 & 96.7/45.6 \\
\bottomrule
\end{tabular}
\end{FitToWidth}  

\label{tab:rabc_data}
  \vspace{-0.3cm}
\end{table}

\begin{table}
  \centering
     \caption{Number of samples per class of BabbleCor.}
\setlength{\tabcolsep}{2.0pt}
  \begin{tabular}{c|ccccc|c}
    \toprule
    partition & \texttt{Non-CAN} & \texttt{CAN} & \texttt{LAU} & \texttt{CRY} & \texttt{JUNK} & Total\\
    \midrule
    training & 1324 & 410 & 42 & 223 & 1651 & 3650 \\ 
    development & 1521 & 352 & 37 & 151 & 1242 & 3303 \\
    testing & 1247 & 545 & 55 & 235 & 1252 & 3334\\
\bottomrule
\end{tabular}
\label{tab:bbc_data}
  \vspace{-0.5cm}
\end{table} 

\noindent\textbf{BabbleCor}~\cite{cychosz2021vocal} contains 11k short audio clips of 52 healthy children (2-36 months) without speech delay. 
The audio clips were collected from day-long recordings of children at home using LENA~\cite{gilkerson2008lena}, an infant-wearable audio recording device. 
Annotators label the clips as one of \texttt{CANONICAL} (containing a consonant to vowel transition
), \texttt{NON-CANONICAL} (not containing a consonant to vowel transition
), \texttt{CRY}, \texttt{LAU}, or \texttt{JUNK} (junk, same as silent or non-speech). The current BabbleCor distribution includes about 91\% of the data used for the Baby Sounds (BS) sub-challenge in the 2019 Interspeech Paralinguistic Challenge~\cite{schuller2019interspeech}. We use the speaker split of the BS challenge. Table~\ref{tab:bbc_data} presents details of our data partition. The mean duration of clips is 0.36s $\pm$ 0.08s std.

\noindent\textbf{My Science Tutor (MyST)}~\cite{ward2011my} contains conversational speech of 1371 students between third and fifth grades with a virtual tutor. 
Our phone recognizer is trained and tested using 
short transcribed utterances ($<15$s) 
to maximize alignment with the shorter vocalizations produced by younger children. We have 84.2h/14.1h/15.1h for training/development/testing set respectively. The mean and std of the speech is 6.6s $\pm$ 3.6s.

\noindent\textbf{Providence}~\cite{demuth2006word} contains longitudinal audio and video recordings of six English-speaking children (3 boys: Alex, Ethan, and William, and 3 girls: Lily, Naima, and Violet) from ages 1 to 4 interacting with their mothers at home.
Annotators transcribed children's speech using SAMPA phonetic symbols. We manually filter out some of the highly noisy recordings and use transcribed child audio for fine-tuning W2V2 PR. In total, we train on 84.0h utterances of four children (Ethan, William, Lily, and Naima) and test on 24.8h utterances of the other two (Alex and Violet). 
The mean and std of the utterances is 3.3s $\pm$ 2.1s.

\begin{figure*}[t]
  \centering
  \includegraphics[width=1.0\linewidth]{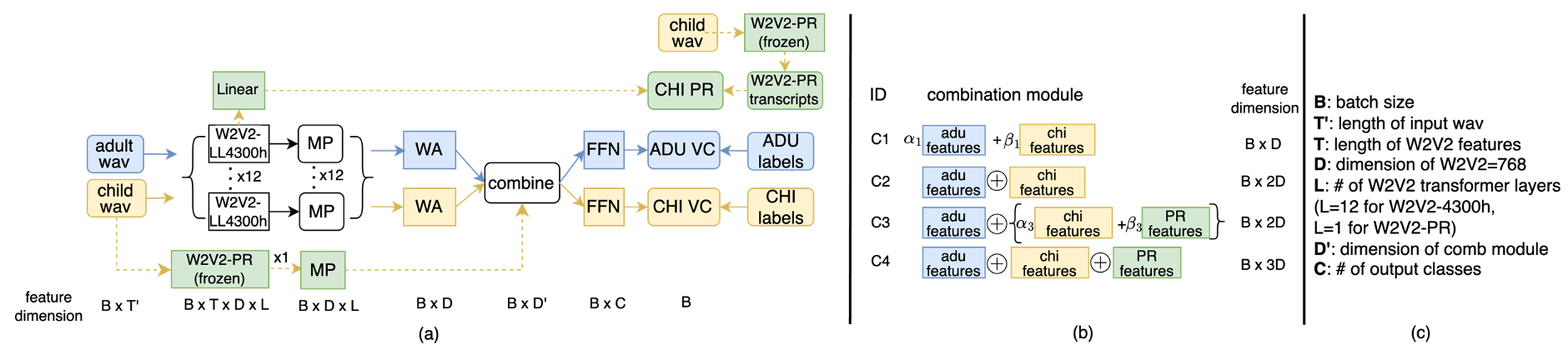}
    \vspace{-0.8cm}
\caption{(a): W2V2 model architecture combining adult audio \& child audio with/without auxiliary W2V2-Children's PR (\textit{W2V2-PR}) features or auxiliary W2V2-Children's PR task. Linear=a linear layer, MP=mean pooling, WA=weighted average, and FFN=feed-forward network. \texttt{ADU VC}, \texttt{CHI VC}, \texttt{CHI PR} denote adult VC, child VC, and auxiliary child PR tiers respectively. 
(b): Illustration of four combination modules. Symbol ``$+$'' means summation and ``$\bigoplus$'' means concatenation. For combination $C1$ and $C3$, $\alpha_i+\beta_i=1$ for $i\in\{1,3\}$. 
(c): Explanation of feature dimension letters.}
  \label{fig:model_combine}
  \vspace{-0.5cm}
\end{figure*}

\section{Methodology}
\subsection{Baseline W2V2 systems for child-adult SD and VC}
Detailed model architecture and training procedures of W2V2 are described in~\cite{wav2vec}. Briefly, W2V2 encodes raw audio into latent embeddings and learns to predict masked embeddings of quantized speech units from contextual embeddings during pretraining. 
We use \textbf{\textit{W2V2-base}} (hidden feature size 768 with 12 transformer layers) in our study.
For the baseline W2V2-based model on RABC, we adapt the model from our previous work~\cite{li2023towards}. Specifically, input waveforms from adult and child audio channels are separately fed into the W2V2 model to extract hidden features from 12 transformer layers. Mean pooling across the duration of the utterance is followed by a weighted average across layers. Two feed-forward networks are used as output tiers for classifying adult and child audio as silence, or one of their VC types. Cross-entropy losses of two output tiers are averaged as the overall loss. We test separate vs.~joint learning of the two audio channels, fine-tuning either \textbf{\textit{W2V2-LL4300h}} (pre-trained on 4300h daylong home recordings of children under 5 years old~\cite{li2023towards}) or \textit{W2V2-base} (pre-trained on 960h of unlabeled adult audio~\cite{wav2vec}). 
Figure~\ref{fig:model_combine} shows baseline joint learning using \textit{W2V2-LL4300h} plus the combination module and auxiliary PR task described in Section~\ref{sec:combining_model}. 

\subsection{Energy thresholding on two audio channels for SD}

For RABC, we also test two energy thresholding (ET) baselines for SD. Each baseline finds thresholds of the child and adult microphones using the training set of each fold, labels vocalization when the energy spectrum is above threshold in a 0.1s frame, and smooths the results with an 11-frame median filter.

\noindent\textbf{Unsupervised ET}: For each audio stream, we test multiple ETs and find the one that best matches the Pydub\footnote{https://github.com/jiaaro/pydub} voice activity detector.  

\noindent\textbf{Weak supervised ET}: Given labels, for each audio stream, we find an ET above which the foreground speaker is always louder than the background speaker.

\subsection{Learning phonetic embeddings using W2V2}

To train phone recognition for children under 4 years old, we follow two-level fine-tuning to gradually reduce age mismatch from adult to child voices, similarly as described in our previous work~\cite{li2024analysis}. We briefly summarize the training procedures as follows. We first convert transcripts of MyST and Providence into international phonetic alphabet (IPA) format. For MyST, we use eng\_to\_ipa software\footnote{https://pypi.org/project/eng-to-ipa/}. For Providence, we directly map SAMPA symbols to IPA. In total, we have 53 unique IPA phones. 


\noindent\textbf{\textit{W2V2-MyST}} We build a CTC-based~\cite{graves2006connectionist}
phone recognizer using one linear layer (hidden dimension 384), Leaky Relu activation, and softmax on top of \textbf{\textit{W2V2-Libri960h}} (\textit{W2V2-base} fine-tuned on 960h of adult LibriSpeech~\cite{librispeech} data). We fine-tune \textit{W2V2-Libri960h} using MyST data to minimize phone-level CTC loss for 40 epochs. For inference, we use CTC greedy decoding without a language model, obtaining 11.2\% phone error rate (PER) on the test set. 

\noindent\textbf{\textit{W2V2-Pro}} We follow similar training and inference setup to fine-tune \textit{W2V2-MyST} using Providence for 5 epochs. We obtain PER of 61.3\% on the test set. 
Note that higher PER does not imply lower utility: Providence, RABC, and BabbleCor children are all $\le 4$ years old while MyST children are older.

\vspace{-0.2cm}
\subsection{Auxiliary W2V2 children's phonetics} 
\label{sec:combining_model}

\vspace{-0.1cm}
We explore three methods to incorporate phonetic information from \textit{W2V2-PR} ({\textit{W2V2-MyST} \& \textit{W2V2-Pro}) into VC: (1) using W2V2 internal representations as an auxiliary \textit{input} features, (2) recognizing pseudo W2V2 phonetic transcripts as an auxiliary \textit{output} task, and (3) combine (1) and (2).  

\noindent\textbf{Auxiliary input features} To leverage background speech information in RABC, we introduce a combination (comb) module to fuse utterance-level W2V2 features from both audio channels via summation or concatenation.  Figure~\ref{fig:model_combine} presents the detailed model architecture. For summation (comb $C1$), we set $\alpha_{1}=0.8$ and $\beta_{1}=0.2$ for \texttt{ADU} tier, and $\alpha_{1}=0.2$ and $\beta_{1}=0.8$ for \texttt{CHI} tier. Therefore, for each tier, features of foreground speaker have a larger weight than background speaker. We also explore combining \textit{W2V2-PR} features with \textit{W2V2-LL4300h} features in the child channel (comb $C3$ and $C4$). During fine-tuning of \textit{W2V2-LL4300h}, we freeze \textit{W2V2-PR} and only extract the last transformer layer features as an auxiliary input. The last transformer layer may encode richer phonetic information as it's closest to the output layer. For both summation modules (comb $C1$ and $C3$), we ensure the weights of both input features sum up to 1. 

\noindent\textbf{Auxiliary output task} We first feed RABC or BabbleCor audio as the input into either W2V2-MyST or W2V2-Pro to generate phonetic pseudo-reference transcripts. 
To encourage \textit{W2V2-LL4300h} to learn children's phone embeddings, we then add phone recognition as an auxiliary output task to the VC system shown in Figure~\ref{fig:model_combine}. Specifically, since \textit{W2V2-LL4300h} best captures phonetic information in its middle transformer layers~\cite{li2024analysis}, we add a linear layer on top of a middle transformer layer and train it using CTC to generate hypothesis transcripts that match the pseudo-reference transcripts with minimum cross-entropy. 

\subsection{Experimental Setup}
All waveforms are downsampled to 16 kHz. For RABC, each fine-tuning experiment is trained 10 epochs with batch size 32 on an NVIDIA 1080 Ti for about 10 hours. 
Adam optimizer sets learning rates (LR) of 1e-4 for the output linear layer and 1e-5 for the W2V2; scheduler with the new-bob technique adjusts LR based on the test set performance (average of unweighted F1-scores over two channels) after each epoch. 
Best performance of the test set is used for each fold. 
SD outputs are computed 
by merging non-silent VC outputs then applying an 11-frame median filter.
SD task is evaluated using diarization error rate (DER). We follow a common standard with 0.25s collar forgiveness over reference segments. We diarize from scratch without knowing the reference voiced segments. Overlapped speech segments are included for evaluating DER. VC task is evaluated using unweighted F1-scores for each speaker over all classes. 
We use a similar setup on BabbleCor by ignoring the adult audio channel. We tune LR as 3e-5/1e-5 for the output linear layer/W2V2 respectively.
Optimal results for summation (comb $C3$) are obtained using $\alpha_3=0.8$ and $\beta_3=0.2$.
We evaluate children's VC using unweighted average recall (UAR), the same metric used in the BS challenge, and unweighted F1-score over all classes.  We also compute 95\% confidence interval on the test set using~\cite{confidence_interval}.
We implement all experiments using SpeechBrain~\cite{speechbrain}. Code and model weights are available
\footnote{https://huggingface.co/lijialudew/}. 
\vspace{-0.3cm}
\section{Results}
\vspace{-0.2cm}
\subsection{Baseline models}
Table~\ref{tab:rabc_baseline_results} summarizes the baseline results. Weak supervised ET improves over unsupervised ET by correcting errors related to noises and background speech ($\mathtt{B_1}$ vs. $\mathtt{B_2}$). Jointly fine-tuning on \textit{W2V2-LL4300h} greatly improves the performance of \textit{W2V2-base} ($\mathtt{B_3}$ vs. $\mathtt{B_4}$) because \textit{W2V2-base} is pre-trained on adult speech only and fails to capture children's phonetics. Separately learning two audio streams yields slight benefit for SD task but not for VC task ($\mathtt{B_4}$ vs. $\mathtt{B_5}$). Thus, we use joint modeling on \textit{W2V2-LL4300h} for the rest of our experiments.

\begin{table}[]
   \caption{Mean and Std of DER and F1 scores of \texttt{ADU} and \texttt{CHI} VC tasks, in percent, over 3-fold cross-validation on RABC corpus trained on baseline models. ET=energy thresholding. Separate W2V2=W2V2-base used for \texttt{ADU} and W2V2-LL4300h used for \texttt{CHI} respectively. }
  \centering
\setlength{\tabcolsep}{1.0pt}
\begin{FitToWidth}[\columnwidth]
  \begin{tabular}{c|c|c|c|c}
    \toprule
    Exp ID & Method & DER & \texttt{ADU}-F1 & \texttt{CHI}-F1 \\
    \midrule
    $\mathtt{B_1}$ & ET  (unsupervised) & 64.4 $\pm$ 6.3 & - &- \\ 
    $\mathtt{B_2}$ & ET (weak supervised)  & 41.4 $\pm$ 4.0 & - &- \\ 
    \midrule
    $\mathtt{B_3}$ & joint (W2V2-base) & 26.7 $\pm$ 15.1 & 81.7 $\pm$ 0.9 & 46.8 $\pm$ 3.5\\
    $\mathtt{B_4}$ & joint (W2V2-LL4300h) & 19.3 $\pm$ 5.1 & \textbf{83.3 $\pm$ 0.2} & \textbf{55.7 $\pm$ 1.7}\\
    \midrule
    $\mathtt{B_5}$ & separate W2V2 & \textbf{19.2 $\pm$ 4.3} & 82.6 $\pm$ 0.1 & 54.4 $\pm$ 1.7\\
\bottomrule
\end{tabular}
\end{FitToWidth}
\label{tab:rabc_baseline_results}
 \smallskip 
\caption{Mean and Std of DER and F1 scores of \texttt{ADU} and \texttt{CHI} VC tasks on RABC corpus trained with different feature inputs ($E_1-E_7$), auxiliary output tasks ($E_8-E_{10}$), or both ($E_{11}$).} 
\setlength{\tabcolsep}{1.0pt}
\begin{FitToWidth}[\columnwidth]
  \begin{tabular}{c|c|c|c|c|c}
    \toprule
    Exp & comb ID & model type &DER & \texttt{ADU}-F1 & \texttt{CHI}-F1 \\
    \midrule
    $\mathtt{E_1}$ & $C1$ & - & {18.7 $\pm$ 4.8} & 83.5 $\pm$ 2.1 & 55.1 $\pm$ 1.3\\
    $\mathtt{E_2}$ & $C2$ & - & 17.3 $\pm$ 5.0 & 83.5 $\pm$ 1.5 & 56.7 $\pm$ 0.6\\
    \midrule\midrule
    $\mathtt{E_3}$ & $C3$  ($\beta_3$=0.5) & W2V2-Pro& {18.1 $\pm$ 5.0} & 83.5 $\pm$ 0.9 & 58.2 $\pm$ 1.2\\
    $\mathtt{E_4}$ & $C4$ & W2V2-Pro & {18.1 $\pm$ 5.2} & 83.4 $\pm$ 1.1 & 57.3 $\pm$ 1.1\\
    \midrule
    $\mathtt{E_5}$ & $C3$ ($\beta_3$=0.5)& W2V2-MyST & {17.8 $\pm$ 5.0} & 83.1 $\pm$ 1.1 & 57.6 $\pm$ 0.1\\
    $\mathtt{E_6}$ & $C3$ ($\beta_3$=0.2)& W2V2-Pro  &{17.8 $\pm$ 3.7} & 83.4 $\pm$ 1.3 & 57.1 $\pm$ 1.1\\
    $\mathtt{E_7}$ & $C3$ ($\beta_3$=0.8)  & W2V2-Pro & {18.3 $\pm$ 4.6} & 82.5 $\pm$ 2.2 & \textbf{58.4 $\pm$ 0.0}\\
    \midrule
    \midrule
    $\mathtt{E_8}$ & $C2$ & W2V2-Pro (PR @ layer 6) & {17.9 $\pm$ 4.6} & \textbf{84.0 $\pm$ 1.7} & 57.9 $\pm$ 1.3\\
    $\mathtt{E_9}$ & $C2$ & W2V2-Pro (PR @ layer 8) &\textbf{17.2 $\pm$ 4.2} & 83.8 $\pm$ 1.7 & 57.9 $\pm$ 1.5\\
    $\mathtt{E_{10}}$ & $C2$ & W2V2-MyST (PR @ layer 8) & {18.8 $\pm$ 3.3} & 83.6 $\pm$ 0.7 & 57.6 $\pm$ 1.1\\
    \midrule\midrule
    $\mathtt{E_{11}}$ & $C3$ ($\beta_3$=0.5) & W2V2-Pro (PR @ layer 8) & {18.1 $\pm$ 3.9} & 82.7 $\pm$ 1.9 & 57.5 $\pm$ 0.6\\

\bottomrule
\end{tabular}
\end{FitToWidth}
\label{tab:rabc_combine_results}

\smallskip
\caption{Sample hypothesis and pseudo reference phonetic transcripts for RABC generated from \textit{W2V2-Pro} ($\mathtt{E_9}$) and \textit{W2V2-MyST} ($\mathtt{E_{10}}$). Top/Bottom two rows show four sample utterances for \texttt{VERB}/\texttt{VOC} respectively. ``-'' indicates insertion/deletion errors.}
\resizebox{0.8\columnwidth}{!}{%
\begin{tabular}{l|ll|ll}
\toprule
    & W2V2-Pro & W2V2-MyST                        & W2V2-Pro      & W2V2-MyST           \\ \midrule
ref & n o n o  & n o\textupsilon \ n o\textupsilon                        & b \textscripta \ \textsci \ b \textupsilon \ k   & w \textscripta \ t e\textsci \ w \textupsilon \ k      \\ 
hyp & n - n -  & n o\textupsilon \ n o\textupsilon & b \textschwa \ \textschwa \ b - -   & -  -  -  -  -  -  - \\ \midrule
ref & m i o  &  m i \textschwa \ r & \textsci \ s i b l u  & \textsci \ \textsci \ z b l u\\ 
hyp & m i -  & m i - - & \textsci \ s i - - - & \textsci \ s s i - -\\ \midrule\midrule
ref & k \ae \ t\textesh \ i & k \textschwa                              & b i f a\textsci \ - s  & g r i f i           \\ 
hyp & - \ae \ t\textesh \ - & - -                              & r i  t k  r s & r r - - -           \\ \midrule
ref & b \ae \ k & \textipa{\dh} \textschwa \ t& m  & \textschwa \ m a\textsci \ n\\ 
hyp & b \ae \ k & - \textschwa \ - & - &  - - a\textsci \ - \\
\bottomrule
\end{tabular}}%
\label{tab:sample_utt}

\bigskip
\caption{UAR and F1 scores for development and testing set of BabbleCor in past studies (top 4 rows) and our studies ($\mathtt{BC_0}$-$\mathtt{BC_3}$), and 95\% Confidence intervals are included for the testing set in parenthesis. Our experiments lack 9\% of the data from the BS challenge. }
\setlength{\tabcolsep}{1.0pt}
\begin{FitToWidth}[\columnwidth]
  \begin{tabular}{c|c|c|c|c|c}
    \toprule
    Exp & Method & Dev-UAR & Dev-F1 & Test-UAR & Test-F1 \\
    \midrule
    & {ComPare2019 baseline}~\cite{schuller2019interspeech} & 54.0 & - & 58.7 & -\\
    &Gosztolya~\cite{gosztolya2019using} & 58.7 & - 
    & 59.5 & - \\
    & Heysem Kaya et al.~\cite{kaya2020combining}& 60.1 & - & 61.4 & -\\
    & Sung-Lin Yeh et al.~\cite{yeh2019using}& 61.3 & - & 62.4 & -\\

    \midrule\midrule
    $\mathtt{BC_0}$ & Fine-tune W2V2-LL4300h & 67.6 & 64.1 & 62.9 (57.9, 66.3) & 64.7(58.7, 68.1)\\
    \midrule
    $\mathtt{BC_1}$ & {$\bigoplus$ W2V2-Pro } & 66.7 & \textbf{64.3} & 63.2 (57.6, 68.1) & 65.0 (58.4, 69.1)\\ 
   $\mathtt{BC_2}$ & {$+$ W2V2-Pro ($\beta_3$=0.2)} & \textbf{70.4} & 64.1 & 62.2 (55.9, 66.5) & 64.5 (57.6, 68.5)\\ 
  $\mathtt{BC_3}$ & { W2V2-Pro (PR @ layer 8)} & 68.9 & {63.4} & \textbf{64.6} (58.5, 70.4)& \textbf{66.0} (59.5, 71.2)\\ 
\bottomrule
\end{tabular}
\end{FitToWidth}
\label{tab:babblecor_results}
\end{table}

\vspace{-0.2cm}
\subsection{Auxiliary W2V2 phonetic information}
Table~\ref{tab:rabc_combine_results} shows the relevant results of combining two audio channels. We observe combining both audio channels ($\mathtt{E_1}$ and $\mathtt{E_2}$) improves DER over 
all baselines ($\mathtt{B_1}-\mathtt{B_5}$), and using concatenation module $\mathtt{E_2}$ yields superior performance.

\noindent\textbf{Analysis of auxiliary input features} By introducing \textit{W2V2-Pro} features, $\mathtt{E_3}$ system consistently improves children's VC over each of the 3 folds (not present in Table~\ref{tab:rabc_combine_results} for brevity) while maintaining a comparable DER. This illustrates the advantages of using children's phonetics to improve children's VC task. 
We perform ablation studies based on comb $C3$ by replacing \textit{W2V2-Pro} with \textit{W2V2-MyST} ($\mathtt{E_5}$), reducing ($\mathtt{E_6}$) or increasing ($\mathtt{E_7}$) the weights of \textit{W2V2-Pro} features. 
Children in the RABC test corpus are closer in age to the 6 Providence children than the 1371 MyST children, which may explain superior performance of \textit{W2V2-Pro} over \textit{W2V2-MyST}.
We also find increasing the weights of \textit{W2V2-Pro} features leads to optimal performance of child VC but slightly hurts adult VC. Perhaps overweighting children's phonetic features may add variability to adult speech embeddings and thus hurt adult VC performance.

\noindent\textbf{Analysis of auxiliary output task} 
Recognizing \textit{W2V2-Pro} pseudo transcripts at layer 6 ($\mathtt{E_8}$) or 8 ($\mathtt{E_9}$), along with comb $C2$, 
achieves our best reported DER and child VC.
Recognizing \textit{W2V2-MyST} pseudo transcripts also helps improve child VC but underperforms \textit{W2V2-Pro} ($\mathtt{E_{10}}$ vs. $\mathtt{E_9}$). We further explore incorporating both phonetic features and pseudo phonetic transcripts ($\mathtt{E_{11}}$). The mean and std of PER of recognizing pseudo transcripts for four systems ($\mathtt{E_8}$-$\mathtt{E_{11}}$) are 74.4\% $\pm$ 3.7\%, 78.6\% $\pm$ 5.4\%,  77.0\% $\pm$ 7.0\%, and 72.2\% $\pm$ 3.4\%, respectively.  Though $E_{11}$ has the lowest PER, it doesn't have the best child VC, thus PER is not a perfect predictor of VC.

\noindent\textbf{Sample recognized phone transcripts}
Table~\ref{tab:sample_utt} shows sample pseudo-reference transcripts generated by the \textit{W2V2-MyST} and \textit{W2V2-Pro} phone recognizers, and the corresponding hypothesis transcripts generated by the multi-task trained vocalization classifier.  All are consistent for \texttt{VERB} segments, i.e., words such as ``no no,'' ``buy book,'' and ``is blue.''  Transcripts of \texttt{VOC} nonsense segments are less consistently transcribed; the pseudo-reference transcripts produced for the same segment by \textit{W2V2-Pro} and \textit{W2V2-MyST} differ significantly, and only the multi-task learner trained by \textit{W2V2-Pro} is able to sometimes generate a hypothesis to match the pseudo-reference. 

\vspace{-0.2cm}
\subsection{BabbleCor}
\vspace{-0.1cm}
Table~\ref{tab:babblecor_results} shows the results for BabbleCor. Fine-tuning \textit{W2V2-LL4300h} yields a competitive baseline ($\mathtt{BC_0}$), compared with previous works. 
Interpolation with \textit{W2V2-Pro} features gives the best UAR on the dev set ($\mathtt{BC_2}$), while adding an auxiliary output task gives the best results on the test set ($\mathtt{BC_3}$). Combining auxiliary phonetic features and the auxiliary output task doesn't improve performance, so we didn't include the relevant results in the Table. 
Possibly, BabbleCor's clips may be too short for \textit{W2V2-Pro} to extract useful phonetic cues from internal representation, but recognizing simple generated pseudo phonetic transcripts is beneficial.
Though the distributed BabbleCor does not contain all data from the BS challenge, results suggest that 
our proposed method achieves or surpasses state-of-the-art performance with almost identical data partition. 

\vspace{-0.3cm}
\section{Conclusion \& Future Work}
\vspace{-0.1cm}
The use of W2V2 features, pre-trained using 4300h of home recordings, improves child-adult SD and VC on two-channel audio.
By using children's phonetic embeddings as auxiliary features, or recognizing children's pseudo-reference phonetic transcripts as an auxiliary task, we improve our proposed method on two corpora with different lengths of child utterances. 
One limitation of our study is that, for privacy reasons, the RABC corpus does not specify each child's diagnosis (autism vs. non-autism), so it is not possible to measure the quality of results across diagnostic categories.
In the future, we aim to extend our approach to autistic children's vocalizations when relevant data are available for research.

\vspace{0.2cm}
\noindent\textbf{Acknowledgement}: This work has been funded by the Jump ARCHES endowment through the Health Care Engineering Systems Center at UIUC.
\newpage
\bibliographystyle{IEEEtran}
\bibliography{mybib}

\begin{thebibliography}{10}
\providecommand{\url}[1]{#1}
\csname url@samestyle\endcsname
\providecommand{\newblock}{\relax}
\providecommand{\bibinfo}[2]{#2}
\providecommand{\BIBentrySTDinterwordspacing}{\spaceskip=0pt\relax}
\providecommand{\BIBentryALTinterwordstretchfactor}{4}
\providecommand{\BIBentryALTinterwordspacing}{\spaceskip=\fontdimen2\font plus
\BIBentryALTinterwordstretchfactor\fontdimen3\font minus \fontdimen4\font\relax}
\providecommand{\BIBforeignlanguage}[2]{{%
\expandafter\ifx\csname l@#1\endcsname\relax
\typeout{** WARNING: IEEEtran.bst: No hyphenation pattern has been}%
\typeout{** loaded for the language `#1'. Using the pattern for}%
\typeout{** the default language instead.}%
\else
\language=\csname l@#1\endcsname
\fi
#2}}
\providecommand{\BIBdecl}{\relax}
\BIBdecl

\bibitem{hodges2020autism}
H.~Hodges, C.~Fealko, and N.~Soares, ``Autism spectrum disorder: definition, epidemiology, causes, and clinical evaluation,'' \emph{Translational pediatrics}, vol.~9, no. Suppl 1, p. S55, 2020.

\bibitem{maenner2023prevalence}
M.~J. Maenner, Z.~Warren, A.~R. Williams, E.~Amoakohene, A.~V. Bakian, D.~A. Bilder, M.~S. Durkin, R.~T. Fitzgerald, S.~M. Furnier, M.~M. Hughes \emph{et~al.}, ``Prevalence and characteristics of autism spectrum disorder among children aged 8 years—autism and developmental disabilities monitoring network, 11 sites, united states, 2020,'' \emph{MMWR Surveillance Summaries}, 2023.

\bibitem{fernell2013early}
M.~A.~E. Elisabeth~Fernell and C.~Gillberg, ``Early diagnosis of autism and impact on prognosis: a narrative review,'' \emph{Clinical Epidemiology}, vol.~5, pp. 33--43, 2013.

\bibitem{gordon2016whittling}
E.~Gordon-Lipkin, J.~Foster, and G.~Peacock, ``Whittling down the wait time: exploring models to minimize the delay from initial concern to diagnosis and treatment of autism spectrum disorder,'' \emph{Pediatric Clinics}, 2016.

\bibitem{lord2000autism}
C.~Lord, S.~Risi, L.~Lambrecht, E.~H. Cook, B.~L. Leventhal, P.~C. DiLavore, A.~Pickles, and M.~Rutter, ``The autism diagnostic observation schedule—generic: A standard measure of social and communication deficits associated with the spectrum of autism,'' \emph{Journal of autism and developmental disorders}, vol.~30, pp. 205--223, 2000.

\bibitem{9053251}
N.~R. Koluguri, M.~Kumar, S.~H. Kim, C.~Lord, and S.~Narayanan, ``Meta-learning for robust child-adult classification from speech,'' in \emph{ICASSP}, 2020, pp. 8094--8098.

\bibitem{krishnamachari2021developing}
S.~Krishnamachari, M.~Kumar, S.~H. Kim, C.~Lord, and S.~Narayanan, ``Developing neural representations for robust child-adult diarization,'' in \emph{IEEE Spoken Language Technology Workshop (SLT)}.\hskip 1em plus 0.5em minus 0.4em\relax IEEE, 2021.

\bibitem{ryant2020third}
N.~Ryant, P.~Singh, V.~Krishnamohan, R.~Varma, K.~Church, C.~Cieri, J.~Du, S.~Ganapathy, and M.~Liberman, ``The third dihard diarization challenge,'' \emph{arXiv preprint arXiv:2012.01477}, 2020.

\bibitem{ozonoff2010prospective}
S.~Ozonoff, A.-M. Iosif, F.~Baguio, I.~C. Cook, M.~M. Hill, T.~Hutman, S.~J. Rogers, A.~Rozga, S.~Sangha, M.~Sigman \emph{et~al.}, ``A prospective study of the emergence of early behavioral signs of autism,'' \emph{Journal of the American Academy of Child \& Adolescent Psychiatry}, 2010.

\bibitem{lyakso2016comparison}
E.~Lyakso, O.~Frolova, and A.~Grigorev, ``A comparison of acoustic features of speech of typically developing children and children with autism spectrum disorders,'' in \emph{SPECOM}, Budapest, Hungary, August 2016, pp. 43--50.

\bibitem{mohanta2022analysis}
A.~Mohanta and V.~K. Mittal, ``Analysis and classification of speech sounds of children with autism spectrum disorder using acoustic features,'' \emph{Computer Speech \& Language}, vol.~72, p. 101287, 2022.

\bibitem{cho2019automatic}
S.~Cho, M.~Liberman, N.~Ryant, M.~Cola, R.~T. Schultz, and J.~Parish-Morris, ``Automatic detection of autism spectrum disorder in children using acoustic and text features from brief natural conversations.'' in \emph{Interspeech}, 2019, pp. 2513--2517.

\bibitem{santos2013very}
J.~F. Santos, N.~Brosh, T.~H. Falk, L.~Zwaigenbaum, S.~E. Bryson, W.~Roberts, I.~M. Smith, P.~Szatmari, and J.~A. Brian, ``Very early detection of autism spectrum disorders based on acoustic analysis of pre-verbal vocalizations of 18-month old toddlers,'' in \emph{ICASSP}.\hskip 1em plus 0.5em minus 0.4em\relax IEEE, 2013, pp. 7567--7571.

\bibitem{baird2017automatic}
A.~Baird, S.~Amiriparian, N.~Cummins, A.~M. Alcorn, A.~Batliner, S.~Pugachevskiy, M.~Freitag, M.~Gerczuk, and B.~Schuller, ``Automatic classification of autistic child vocalisations: A novel database and results,'' in \emph{Interspeech}, 2017.

\bibitem{cheng2023computer}
M.~Cheng, Y.~Zhang, Y.~X. andYueran Pan, X.~Li, W.~Liu, C.~Yu, D.~Zhang, Y.~Xing, X.~Huang, F.~Wang, C.~You, Y.~Zou, Y.~Liu, F.~Liang, H.~Zhu, C.~Tang, H.~Deng, X.~Zou, and M.~Li, ``Computer-aided autism spectrum disorder diagnosis with behavior signal processing,'' \emph{IEEE Transactions on Affective Computing}, vol.~14, no.~4, pp. 2982--3000, 2023.

\bibitem{8443030}
M.~Kokot, F.~Petric, M.~Cepanec, D.~Miklić, I.~Bejić, and Z.~Kovačić, ``Classification of child vocal behavior for a robot-assisted autism diagnostic protocol,'' in \emph{2018 26th Mediterranean Conference on Control and Automation (MED)}, 2018, pp. 1--6.

\bibitem{wav2vec}
A.~Baevski, Y.~Zhou, A.~Mohamed, and M.~Auli, ``wav2vec 2.0: A framework for self-supervised learning of speech representations,'' in \emph{NeurIPS}, 2020.

\bibitem{xu2023understanding}
A.~Xu, R.~Hebbar, R.~Lahiri, T.~Feng, L.~Butler, L.~Shen, H.~Tager-Flusberg, and S.~Narayanan, ``Understanding spoken language development of children with {ASD} using pre-trained speech embeddings,'' in \emph{Interspeech}, 2023.

\bibitem{lahiri2023robust}
R.~Lahiri, T.~Feng, R.~Hebbar, C.~Lord, S.~H. Kim, and S.~Narayanan, ``Robust self supervised speech embeddings for child-adult classification in interactions involving children with autism,'' in \emph{Interspeech}, 2023.

\bibitem{Rehg_2013_CVPR}
J.~Rehg, G.~Abowd, A.~Rozga, M.~Romero, M.~Clements, S.~Sclaroff, I.~Essa, O.~Ousley, Y.~Li, C.~Kim \emph{et~al.}, ``Decoding children's social behavior,'' in \emph{CVPR}, June 2013.

\bibitem{cychosz2021vocal}
M.~Cychosz, A.~Cristia, E.~Bergelson, M.~Casillas, G.~Baudet, A.~S. Warlaumont, C.~Scaff, L.~Yankowitz, and A.~Seidl, ``Vocal development in a large-scale crosslinguistic corpus,'' \emph{Developmental science}, 2021.

\bibitem{gilkerson2008lena}
J.~Gilkerson and J.~A. Richards, ``The lena natural language study,'' \emph{Boulder, CO: LENA Foundation. Retrieved March}, vol.~3, p. 2009, 2008.

\bibitem{schuller2019interspeech}
B.~Schuller, A.~Batliner, C.~Bergler, F.~B. Pokorny, J.~Krajewski, M.~Cychosz, R.~Vollmann, S.-D. Roelen, S.~Schnieder, E.~Bergelson \emph{et~al.}, ``The interspeech 2019 computational paralinguistics challenge: Styrian dialects, continuous sleepiness, baby sounds \& orca activity,'' in \emph{Interspeech}.\hskip 1em plus 0.5em minus 0.4em\relax ISCA, 2019.

\bibitem{ward2011my}
W.~Ward, R.~Cole, D.~Bolanos, C.~Buchenroth-Martin, E.~Svirsky, S.~V. Vuuren, T.~Weston, J.~Zheng, and L.~Becker, ``My science tutor: A conversational multimedia virtual tutor for elementary school science,'' \emph{ACM Transactions on Speech and Language Processing (TSLP)}, vol.~7, no.~4, pp. 1--29, 2011.

\bibitem{demuth2006word}
K.~Demuth, J.~Culbertson, and J.~Alter, ``Word-minimality, epenthesis and coda licensing in the early acquisition of english,'' \emph{Language and speech}, vol.~49, no.~2, pp. 137--173, 2006.

\bibitem{li2023towards}
J.~Li, M.~Hasegawa-Johnson, and N.~L. McElwain, ``Towards robust family-infant audio analysis based on unsupervised pretraining of wav2vec 2.0 on large-scale unlabeled family audio,'' in \emph{Interspeech}, 2023.

\bibitem{li2024analysis}
------, ``Analysis of self-supervised speech models on children's speech and infant vocalizations,'' in \emph{IEEE Workshop on Self-Supervision in Audio, Speech and Beyond (SASB)}, 2024.

\bibitem{graves2006connectionist}
A.~Graves, S.~Fern{\'a}ndez, F.~Gomez, and J.~Schmidhuber, ``Connectionist temporal classification: labelling unsegmented sequence data with recurrent neural networks,'' in \emph{ICML}, 2006, pp. 369--376.

\bibitem{librispeech}
V.~Panayotov, G.~Chen, D.~Povey, and S.~Khudanpur, ``Librispeech: An asr corpus based on public domain audio books,'' in \emph{ICASSP}, 2015, pp. 5206--5210.

\bibitem{confidence_interval}
\BIBentryALTinterwordspacing
L.~Ferrer and P.~Riera, ``Confidence intervals for evaluation in machine learning,'' 2024. [Online]. Available: \url{https://github.com/luferrer/ConfidenceIntervals}
\BIBentrySTDinterwordspacing

\bibitem{speechbrain}
M.~Ravanelli, T.~Parcollet, P.~Plantinga, A.~Rouhe, S.~Cornell, L.~Lugosch, C.~Subakan, N.~Dawalatabad, A.~Heba, J.~Zhong \emph{et~al.}, ``{SpeechBrain}: A general-purpose speech toolkit,'' 2021, arXiv:2106.04624.

\bibitem{gosztolya2019using}
G.~Gosztolya, ``Using fisher vector and bag-of-audio-words representations to identify styrian dialects, sleepiness, baby \& orca sounds,'' in \emph{Interspeech}, 2019.

\bibitem{kaya2020combining}
H.~Kaya, O.~Verkholyak, M.~Markitantov, and A.~Karpov, ``Combining clustering and functionals based acoustic feature representations for classification of baby sounds,'' in \emph{International Conference on Multimodal Interaction}, 2020, pp. 509--513.

\bibitem{yeh2019using}
S.-L. Yeh, G.-Y. Chao, B.-H. Su, Y.-L. Huang, M.-H. Lin, Y.-C. Tsai, Y.-W. Tai, Z.-C. Lu, C.-Y. Chen, T.-M. Tai \emph{et~al.}, ``Using attention networks and adversarial augmentation for styrian dialect continuous sleepiness and baby sound recognition.'' in \emph{Interspeech}, 2019, pp. 2398--2402.

\end{thebibliography}

\end{document}